\documentclass{pasj01}

\Received{$\langle$reception date$\rangle$}
\Accepted{$\langle$acception date$\rangle$}
\Published{$\langle$publication date$\rangle$}

\usepackage{appendix}
\usepackage{url}
\usepackage{lineno}
\newcommand{\PKHY}[1]{{{#1}}}
\newcommand{\HILI}[1]{{{#1}}}

\begin{document}

\title{Examining a hadronic $\gamma$-ray scenario for the radiative shell \& molecular clouds of the old GeV supernova remnant G298.6$-$0.0}
\author{
Paul K. H. Yeung\altaffilmark{1,2,*},
Shiu-Hang Lee\altaffilmark{3,4},
Tsunefumi Mizuno\altaffilmark{5},
Aya Bamba\altaffilmark{1,6,7},
}
\altaffiltext{1}{Department of Physics, The University of Tokyo, 7-3-1 Hongo, Bunkyo-ku, Tokyo 113-0033, Japan}
\altaffiltext{2}{Institute for Cosmic Ray Research, University of Tokyo, 5-1-5, Kashiwa-no-ha, Kashiwa, Chiba 277-8582, Japan}
\altaffiltext{3}{Department of Astronomy, Kyoto University, Kitashirakawa, Oiwake-cho, Sakyo-ku, Kyoto 606-8502, Japan}
\altaffiltext{4}{Kavli Institute for the Physics and Mathematics of the Universe (WPI), The University of Tokyo, Kashiwa 277-8583, Japan}
\altaffiltext{5}{Hiroshima Astrophysical Science Center, Hiroshima University, 1-3-1 Kagamiyama, Higashi-Hiroshima, Hiroshima 739-8526, Japan}
\altaffiltext{6}{Research Center for the Early Universe, School of Science, The University of Tokyo, 7-3-1 Hongo, Bunkyo-ku, Tokyo 113-0033, Japan}
\altaffiltext{7}{Trans-Scale Quantum Science Institute, The University of Tokyo, Tokyo 113-0033, Japan}
\email{pkh91yg@icrr.u-tokyo.ac.jp}

\KeyWords{cosmic rays --- gamma rays: ISM --- X-rays: ISM --- radio lines: ISM --- ISM: supernova remnants --- ISM: individual objects (G298.6$-$0.0, 4FGL J1213.3$-$6240e)}

\maketitle

\begin{abstract}

Based on the 13.7~yr \emph{Fermi}-LAT data, \citet{Yeung2023} claimed detection of two  $\gamma$-ray sources (namely Src-NE and Src-NW) associated with the supernova remnant (SNR) G298.6$-$0.0, and interpreted it as an old GeV SNR interacting with molecular clouds (MCs). In this follow-up study, we refine the flux measurements below 2~GeV with \emph{Fermi}-LAT event types of better angular reconstruction. Then, we report  our   $\gamma$-ray spectral fittings and cosmic-ray phenomenology in a hadronic scenario, considering both the shell and MC regions of SNR G298.6$-$0.0. We confirm that both the $\gamma$-ray spectra of Src-NE and Src-NW exhibit spectral breaks at $1.50_{-0.50}^{+0.60}$~GeV and $0.68_{-0.11}^{+0.32}$~GeV, respectively.   Src-NW  has a harder broadband photon index than Src-NE, suggesting an appreciable  difference between the physical separations of their respective emission sites from SNR G298.6$-$0.0. The cosmic-ray spectrum responsible for Src-NE starts with a minimum energy $E_\mathrm{CR,min}=1.38_{-0.16}^{+0.47}$~GeV, and has a proton index $\Gamma_\mathrm{CR}=2.57_{-0.21}^{+0.18}$ below the exponential cutoff energy $E_\mathrm{CR,max}=240_{-150}^{+240}$~GeV.   \PKHY{Accordingly, we argue that Src-NE is dominated by the SNR shell, while only a minor  portion of lower-energy emission is contributed by the  MCs interacting with the SNR.}  The cosmic-ray population for Src-NW starts at a higher energy such that the $E_\mathrm{CR,min}$ ratio of Src-NW to Src-NE is $\gtrsim$2. The high $E_\mathrm{CR,min}$, as well as the high cosmic-ray energy density required ($\sim$26~eV~cm$^{-3}$),  supports the interpretation that Src-NW is predominantly the  $\gamma$-ray emission from the farther MCs  being bombarded by protons that had earlier escaped from SNR G298.6$-$0.0.  By comparing the high-energy features of G298.6$-$0.0 with those of analogical SNRs, especially SNR W28 and SNR W44,  we further constrain the age of SNR G298.6$-$0.0 to be 10--30~kyr, \PKHY{under the assumption of a purely hadronic scenario}.

\end{abstract}

\section{Introduction}

GeV--TeV $\gamma$-rays (e.g., \cite{Aharonian2004, Acero2016}) from the regions of supernova remnants (SNRs) provide evidence that SNR shock fronts are powerful sites of cosmic-ray acceleration. When a hadronic cosmic-ray particle (for example, a proton or an atomic nucleus) from an SNR collides with a target hadron, a neutral pion is produced which quickly decays into two $\gamma$-ray photons that we can observe. Such hadronic collisions, as well as the accompanied $\gamma$-ray emissions, could occur at the molecular clouds (MCs) upstream of the SNR shock in addition to the radiative shell downstream of the  shock \citep{Slane2015}. SNR W28, SNR W44, SNR HESS J1731-347, and SNR HB9 are four conspicuous examples where $\gamma$-rays are detected at both post-shock SNR shells and pre-shock MCs \citep{Cui2018, Peron2020, Cui2019, Oka2022}.

\HILI{Thanks to the \emph{Fermi} Large Area Telescope (\emph{Fermi}-LAT) and AGILE, spectral breaks/cutoffs in the GeV $\gamma$-ray band \citep{Abdo2009, Giuliani2011, Acero2016, Ambrogi2019} are} commonly observed for SNR--MC interaction systems, reflecting that the  over-energetic particles easily escape  from the vicinities of SNRs. More intriguingly,   a number of attempts have been made to formulate the relationship between the $\gamma$-ray spectral shape and the SNR age (e.g., \cite{Dermer2013, Bamba2016, Zeng2019, Suzuki2020, Suzuki2022}). All these studies reveal that   an older SNR is generally associated with a softer $\gamma$-ray spectrum   and a lower-energy spectral peak. Such an observed trend is consistent with a theory that the escape of cosmic rays from SNRs gradually develops from highest-energy particles to lowest-energy particles with time \citep{Ptuskin2003, Ptuskin2005}. It is particularly important for us to investigate $\gamma$-rays associated with old ($\gtrsim$10~kyr) SNRs, for the sake of better understanding the later evolution stages of the cosmic-ray escape.

SNR G298.6$-$0.0 was detected by Molonglo at 408~MHz \citep{Shaver1970} and by MOST at 843~MHz \citep{Kesteven1987, Whiteoak1996}, and was found to have a flat radio spectral index of $-$0.3 \citep{Shaver1970}. \citet{Reach2006} claimed a possible detection of infrared emission (possibly induced by shock heating of the interstellar medium) from the direction of G298.6$-$0.0, proposing a  collision of the SNR shock with a high-density medium. The interaction of this SNR with MCs has been confirmed in our previous multiwavelength study, \HILI{which reports a $\gamma$-ray spectral break detected with \emph{Fermi}-LAT and nearby MCs traced by NANTEN CO observations} \citep{Yeung2023}.  To the southwest of G298.6$-$0.0 (about 0.3$^\circ$ apart), there is another SNR -- G298.5$-$0.3, detected by Molonglo and MOST at 408~MHz and 843~MHz, respectively \citep{Shaver1970, Whiteoak1996}. A  possible detection of infrared emission from the direction of G298.5$-$0.3 \citep{Reach2006} makes it likely to be interacting with dense medium as well.

\begin{figure}
    \centering
    \includegraphics[width=80mm]{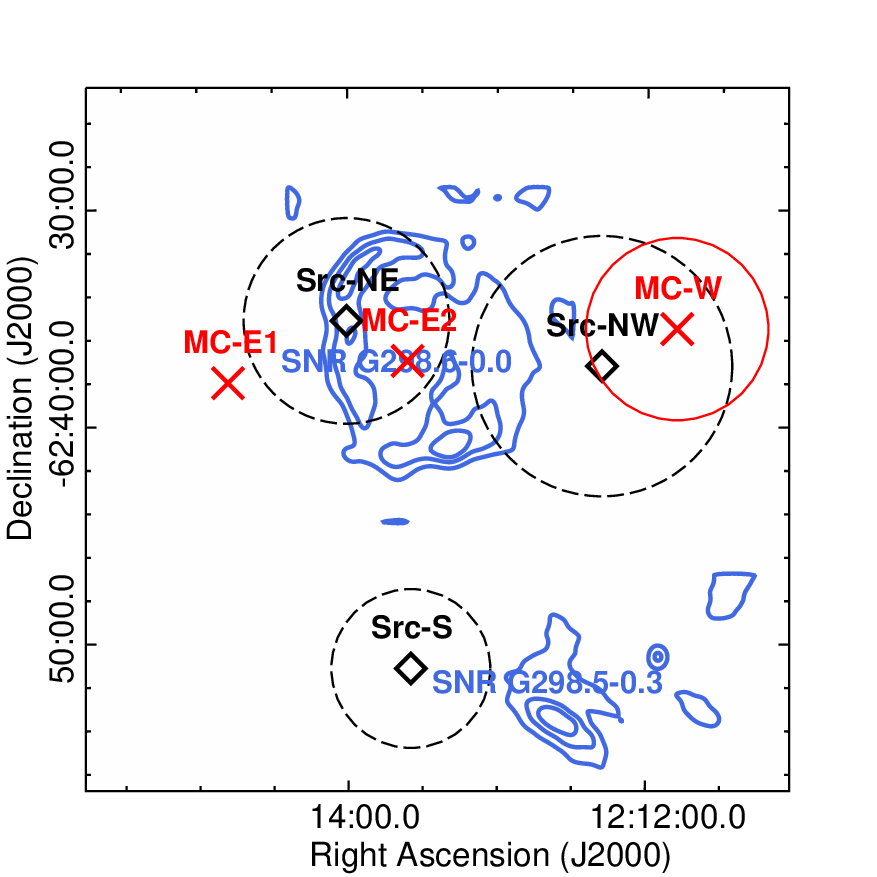}
    \caption{The spatial configuration of the SNRs, MC clumps and $\gamma$-ray components in our targeted region. The MOST 843~MHz radio continuum contours of SNR G298.6$-$0.0 and SNR G298.5$-$0.3 \citep{Whiteoak1996} are plotted in blue. Three clumps of dense MCs at approximately the same distance from us as G298.6$-$0.0 ($\sim$10.1~kpc), which are traced by $^{12}$CO($J$=1--0) (115~GHz) line emission  (Figure~5 of \cite{Yeung2023}), are marked as red crosses. The approximate size of the clump MC-W is indicated by the red circle. The \emph{Fermi}-LAT $\gamma$-ray  centroids  of Src-NE, Src-NW and Src-S are marked as black diamonds encircled by their respective \PKHY{error circles at the 95\% confidence level} (taken from Figure~1 of \cite{Yeung2023}). }
    \label{schematic}
\end{figure}

By analyzing  13.7~yr \emph{Fermi}-LAT $\gamma$-ray data, we  previously identified three spatial components in/around the sky region of these two SNRs: Src-NE, Src-NW and Src-S \citep{Yeung2023}. Src-NE is at the eastern shell of G298.6$-$0.0, Src-NW is adjacent to the western edge of G298.6$-$0.0, and Src-S is adjacent to the eastern edge of the other SNR G298.5$-$0.3. Additionally, MCs traced by $^{12}$CO($J$=1--0) (115~GHz) line emission are present within both the error circles of Src-NE and Src-NW (Figure~5 of \cite{Yeung2023}). These MCs are potentially interacting with SNR G298.6$-$0.0 since they are approximately equidistant (both at a kinematic distance of $\sim$10.1~kpc) from us. The spatial configuration of these SNRs, MC clumps and $\gamma$-ray components is graphically illustrated in Figure~\ref{schematic}.  Considering the $\gamma$-ray spectral break energy of Src-NE in context with the evolution trend of SNRs, we provided the very first estimation of the age of SNR G298.6$-$0.0 ($>$10~kyr; \cite{Yeung2023}). This lower limit on the age is also supported by our non-detection of ionising/recombining plasma's X-ray emission and a large physical radius of $\sim$15.5~pc \citep{Yeung2023}. Such an old age, as well as potential emission from both the shell and MCs, encourages us to further investigate this SNR with a cosmic-ray phenomenological approach.

This follow-up paper reports our \PKHY{improved spectroscopy and cosmic-ray energetic estimates} of the detected $\gamma$-rays from the regions of SNR G298.6$-$0.0, SNR G298.5$-$0.3 and their vicinities.  Comparing the $\gamma$-ray spatial location of Src-NE and Src-NW with the radio continuum morphology of SNR G298.6$-$0.0 and the MC distributions traced by CO line emissions (Figure~\ref{schematic}), we propose this hadronic scenario: Src-NE is the combined emission of the SNR shell and nearer MCs, while Src-NW is the emission from farther MCs. Concerning the origins of Src-S, we consider adjacent SNR G298.5$-$0.3 in addition to SNR G298.6$-$0.0. In spectral fittings, we adopt both the flux measurements of \citet{Yeung2023} and the refined measurements of this work.

\section{Improved reductions \& re-analyses of \emph{Fermi}-LAT $\gamma$-ray data}

In \citet{Yeung2023}, we   reconstructed the spectral energy distribution  for each spatial component (Figure~2 of \cite{Yeung2023}). We alerted about the partial discrepancy between the broken-power-law model (obtained from a broadband fit) and the binned spectrum (obtained from an independent fitting for each energy bin) of Src-NE, which is due to the significant  overlap of the Galactic diffuse component (gll\_iem\_v07.fits) with the point-spread-function (PSF)-convolved model of Src-NE and   the inaccurate energy dependence of the Galactic diffuse component for that region (see \S4.3 of \cite{Yeung2023}). In other words, the fittings for Src-NE suffer from a strong correlation with the Galactic diffuse component.  With regard to this issue, in this work, we refine the flux measurements at lower energies with improved data reductions.

We achieve the improvements by \HILI{adopting PSF1+PSF2+PSF3 data\footnote{Four quartiles of LAT data, in ascending order of the quality of the reconstructed direction, are PSF0, PSF1, PSF2 and PSF3.} instead of FRONT+BACK data\footnote{Photons converted into pairs in the front and back sections of the LAT tracker respectively.}.} This sacrifices about one-fourth of photon statistics in exchange for better angular resolution\footnote{\label{slac}\emph{Fermi} LAT Performance $<$\url{http://www.slac.stanford.edu/exp/glast/groups/canda/lat_Performance.htm}$>$}. Such a cut on the event type is particularly important for lower energies (see Table~1 of \cite{Abdollahi2022}), and hence is suitable for addressing the aforementioned issue.  Because of the reduced PSF size of PSF1+PSF2+PSF3 data, the overlap between the Galactic diffuse component and the PSF-convolved model of Src-NE is also reduced, and the fittings for Src-NE become relatively less sensitive to the inaccurate energy dependence of the Galactic diffuse component. \PKHY{In Appendix~\ref{FTBK_PSF123_2GeV}, we evaluate the sufficiency of the achieved improvement.}

Concerning the other details, we follow the strategy of \citet{Yeung2023}. The  Fermitools version 2.0.8  is  used.  We select  Pass 8 (P8R3) `SOURCE' class events collected between August 4, 2008, and April 28, 2022. The region of interest (ROI) we choose  is $21^{\circ}\times21^{\circ}$ centered at RA =$12^{h}13^{m}40.80^{s}$, Dec=$-62^{\circ}37^{'}12.0^{''}$ (J2000). For each individual energy bin of the revised spectral energy distributions, we perform a  binned maximum-likelihood analysis, with an angular bin size of 0.05$^\circ$. To better model the background, the Galactic diffuse component (gll\_iem\_v07.fits), the isotropic diffuse component (corresponding to the instrument response function P8R3\_SOURCE\_V2), and the sources in the \emph{Fermi} LAT 12-Year Point Source Catalog (4FGL-DR3; \cite{Abdollahi2022}) are included as background sources in our analyses.  We leave the spectral parameters of the sources within 5.5$^\circ$ from the ROI center (including the prefactor and index of the Galactic diffuse background as well as the normalization of the isotropic diffuse background) free to vary in each analysis. For the sources at angular separation beyond 5.5$^\circ$ from the ROI center, their spectral parameters are fixed to the catalog values. Following \citet{Yeung2023}, we replace 4FGL J1213.3$-$6240e (an extended source enclosing both SNR G298.6$-$0.0 and SNR G298.5$-$0.3) with Src-NE, Src-NW and Src-S  that we detected.

\section{Refined results of $\gamma$-ray spectroscopy}
\label{GRspec}

First of all, we perform a maximum-likelihood fit with coherent selection of the PSF1+PSF2+PSF3 data in 0.3--100~GeV. The energy dispersion correction is enabled  for the count spectra of Src-NE, Src-NW, Src-S and all background sources except the isotropic diffuse background, as recommended by the \emph{Fermi} Science Support Center. For each spectrum of our targeted sources, we examine  a power-law (PL) model\footnote{$N_0$ and $\Gamma$ represent the normalisation and photon index, respectively.}:
\begin{equation}
\frac{dN}{dE} = N_0 \left(\frac{E}{3~\mathrm{GeV}}\right)^{-\Gamma},
\end{equation}
and a broken-power-law (BKPL) model\footnote{$E_\mathrm{br}$ represents the energy of the spectral break. $\Gamma_1$ and $\Gamma_2$ represent the photon indices below and above $E_\mathrm{br}$ respectively. }:
\begin{equation}
\frac{dN}{dE} = N_\mathrm{0} \times \left\{
\begin{array}{ll}
	\left(\frac{E}{E_\mathrm{br}}\right)^{-\Gamma_1}\,\mathrm{if}\,E<E_\mathrm{br}  \\
	\left(\frac{E}{E_\mathrm{br}}\right)^{-\Gamma_2}\,\mathrm{if}\,E\geq E_\mathrm{br} 
\end{array}
\right. .
\end{equation}

\PKHY{The 2$\Delta$ln(likelihood) values between BKPL and PL are 28.8, 3.8 and 9.5 for Src-NE, Src-NW and Src-S, respectively. It is worth mentioning that the BKPL fitting for Src-S, contrary to those for Src-NE and Src-NW, gives $\Gamma_1>\Gamma_2$ that is unexpected by a single-origin scenario in the theoretical framework of this paper. Nonetheless, we will evaluate the genuineness of the spectral break for each spatial component more rigorously at the end of this section.}  In the following, we fix all parameters except the normalisations of Src-NE, Src-NW, Src-S, the Galactic diffuse background and the isotropic diffuse background at the values of this global fit. \PKHY{With this setting, we simultaneously iterate their respective fluxes for individual energy bins. The degree of improvement achieved by PSF1+PSF2+PSF3 data is discussed in Appendix~\ref{FTBK_PSF123_2GeV}.}

Our new analysis yields a relatively higher flux and a relatively softer spectrum for the 0.3--2~GeV emission of Src-NE, compared to our previously published results in \citet{Yeung2023}. For Src-NW, its 300--646~MeV fluxes suggest a sharp spectral turnover that was not significantly detected in our previous work.  Our new results for Src-S, similar to our old results in \citet{Yeung2023}, do not seem to reveal a spectral turnover.

Above 2~GeV, we confirm that the flux of each spatial component in each energy bin reconstructed with PSF1+PSF2+PSF3 data is essentially consistent with that reconstructed with FRONT+BACK data (compare the blue bins and red dotted lines of Figures~\ref{SED_Src-NE},~\ref{SED_Src-NW}~and~\ref{SED_Src-S}; \PKHY{a further confirmation is presented in Appendix~\ref{FTBK_PSF123_2GeV}}). More importantly, FRONT+BACK data above 2~GeV has $\sim$33\% increased photon statistics with respect to the PSF1+PSF2+PSF3 data while maintaining sufficiently high angular resolution$^{\ref{slac}}$. Therefore, we continue to use the FRONT+BACK results of \citet{Yeung2023} in this work for the 2--100~GeV bins, but replace the old $<$2~GeV flux points with the PSF1+PSF2+PSF3 results. For all flux points with significant detections, we take into account the systematic uncertainties stemming from the LAT effective area\footnote{Evaluating Aeff Systematics $<$\url{https://fermi.gsfc.nasa.gov/ssc/data/analysis/scitools/Aeff_Systematics.html}$>$} and energy dispersion\footnote{Pass 8 Analysis and Energy Dispersion $<$\url{https://fermi.gsfc.nasa.gov/ssc/data/analysis/documentation/Pass8_edisp_usage.html}$>$}, and we combine the statistical and systematic uncertainties in quadrature.  The revised spectral energy distributions   are demonstrated in Figures~\ref{SED_Src-NE},~\ref{SED_Src-NW}~and~\ref{SED_Src-S}.

\begin{figure}
    \centering
    \includegraphics[width=80mm]{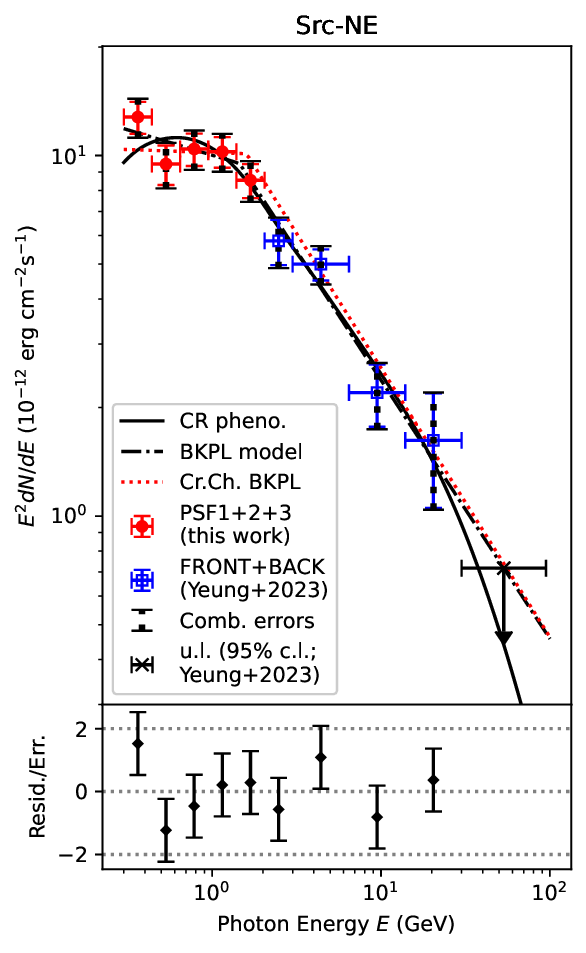}
    \caption{The $\gamma$-ray spectral energy distribution of Src-NE (top), and the residual (data$-$model) divided by the corresponding combined uncertainty when fitting the cosmic-ray phenomenological model to the plotted data (bottom). PSF1+PSF2+PSF3 data below 2~GeV (red filled circles) and FRONT+BACK data above 2~GeV (blue open squares) are adopted for Chi-Square fittings of models (black lines). The combined uncertainties, each of which is the statistical and systematic uncertainties added in quadrature, are plotted in a black-dotted style. Upper limits at the 95\% confidence level are marked as black crosses with a down arrow. The red dotted line is a model for crosschecking, which is reconstructed with a maximum-likelihood fit to the PSF1+PSF2+PSF3 data in 0.3--100~GeV. The grey dotted lines in the bottom panel mark the values of 0 and $\pm$2. A partial duplicate of this figure is presented in Figure~\ref{apx}, supplemented with some crosschecking information.}
    \label{SED_Src-NE}
\end{figure}

\begin{figure}
    \centering
    \includegraphics[width=80mm]{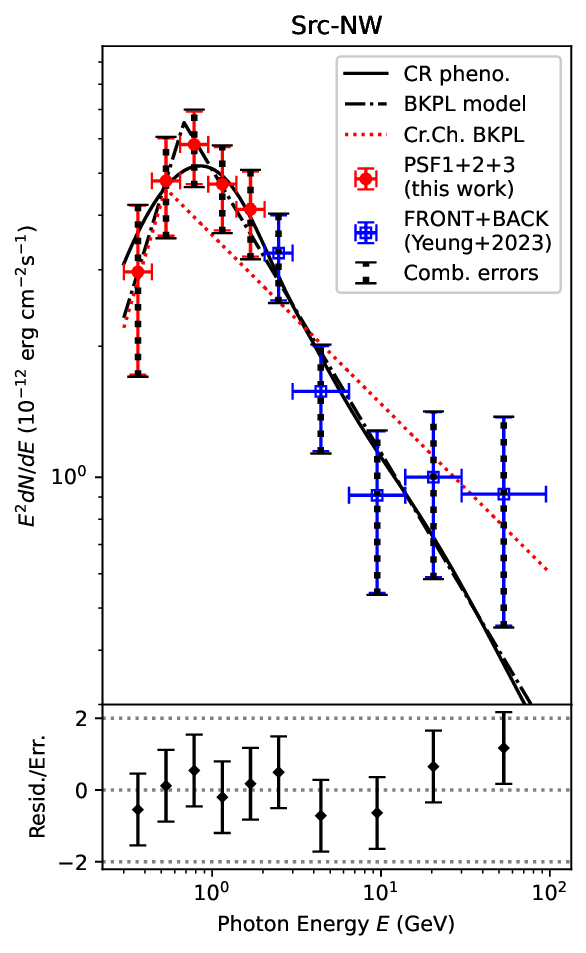}
    \caption{The $\gamma$-ray spectral energy distribution of Src-NW (top), and the residual (data$-$model) divided by the corresponding combined uncertainty when fitting the cosmic-ray phenomenological model to the plotted data (bottom). The colours, symbols and linestyles have the same meanings as in Figure~\ref{SED_Src-NE}. }
    \label{SED_Src-NW}
\end{figure}

\begin{figure}
    \centering
    \includegraphics[width=80mm]{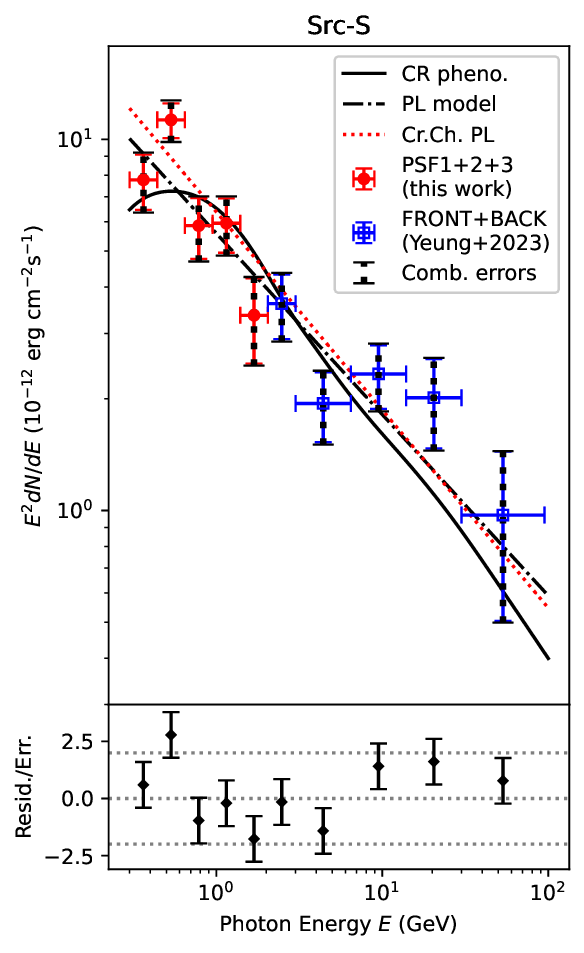}
    \caption{The $\gamma$-ray spectral energy distribution of Src-S (top), and the residual (data$-$model) divided by the corresponding combined uncertainty when fitting the cosmic-ray phenomenological model to the plotted data (bottom). The colours, symbols and linestyles have the same meanings as in Figure~\ref{SED_Src-NE}. }
    \label{SED_Src-S}
\end{figure}

For each spectral energy distribution, we perform Chi-Square fittings with  PL, BKPL, \PKHY{and an exponential-cutoff-power-law (ECPL) model}:
\begin{equation}
\frac{dN}{dE} = N_0 \left(\frac{E}{3~\mathrm{GeV}}\right)^{-\Gamma}\mathrm{exp}(-\frac{E}{E_\mathrm{cut}}).
\end{equation}
The combined uncertainties are adopted in calculating $\chi^2$. We impose an additional constraint on the fitting for Src-NE, forcing the model-predicted flux at $\sim$53~GeV to be below  the upper limit at the 95\% confidence level.  The fitting results are tabulated in Table~\ref{PL_BKPL}.

\begin{table*}
    \caption{$\gamma$-ray  spectral parameters  and statistics for different spatial components, fitted with hybrid selection of PSF1+PSF2+PSF3 data below 2~GeV  and FRONT+BACK data above 2~GeV.}
    \begin{tabular}{lccc}
\hline            & Src-NE             & Src-NW    & Src-S     \\
\hline\multicolumn{4}{c}{PL}            \\
$N_0$ (10$^{-12}$~cm$^{-2}$~s$^{-1}$~GeV$^{-1}$)         &  293 $\pm$ 10   &   157 $\pm$ 15   &   226 $\pm$ 18         \\
$\Gamma$       &  2.62 $\pm$ 0.01   &   2.42 $\pm$ 0.07   &   2.49 $\pm$ 0.06         \\
$\chi^2$/d.o.f.\footnotemark[$*$]  &  26.3 / 7    &    13.5 / 8    &    15.0 / 8         \\
\hline\multicolumn{4}{c}{BKPL}          \\
$N_0$ (10$^{-9}$~cm$^{-2}$~s$^{-1}$~GeV$^{-1}$)         &  2.64 $\pm$ 0.42   &   8.82 $\pm$ 1.68   &   0.08 $\pm$ 0.02         \\
$\Gamma_1$      &  2.14 $\pm$ 0.15   &   0.74 $\pm$ 1.30   &   2.58 $\pm$ 0.08         \\
$E_\mathrm{br}$ (GeV) &  1.50$_{-0.50}^{+0.60}$   &   0.68$_{-0.11}^{+0.32}$   &   4.40 $\pm$ 0.05         \\
$\Gamma_2$      &  2.72$_{-0.03}^{+0.10}$   &   2.65 $\pm$ 0.11   &   2.23 $\pm$ 0.13         \\
$\chi^2$/d.o.f.\footnotemark[$*$]  &  4.3 / 5    &    3.6 / 6    &    11.6 / 6         \\
$\Delta$BIC w.r.t. PL  &       $-$17.6		&	$-$5.2		&	1.2              \\ 
F-test p-value w.r.t. PL  &       0.011	&	0.019	&	0.46              \\ 
2$\Delta$ln(likelihood) w.r.t. PL\footnotemark[$\dag$]  &       28.8	&  3.8	&  9.5              \\  \hline
\multicolumn{4}{c}{ECPL}            \\
$N_0$ (10$^{-12}$~cm$^{-2}$~s$^{-1}$~GeV$^{-1}$)         &  534 $\pm$ 97   &   157 $\pm$ 15   &   226 $\pm$ 18         \\
$\Gamma$       &   2.24 $\pm$ 0.11  &   2.42 $\pm$ 0.07   &   2.49 $\pm$ 0.06         \\
$E_\mathrm{cut}$ (GeV) &   $11.7_{-4.1}^{+9.1}$   & $\infty$($>$1.5)\footnotemark[$\ddag$] & $\infty$($>$72)\footnotemark[$\ddag$]     \\
$\chi^2$/d.o.f.\footnotemark[$*$]  & 7.0  / 6    &    13.5 / 7    &    15.0 / 7         \\
$\Delta$BIC w.r.t. PL  &       $-$17.1		&	2.3		&	2.3              \\ 
F-test p-value w.r.t. PL  &    0.0066   &     1    &        1      \\  \hline
\end{tabular}

    \label{PL_BKPL}
  \begin{tabnote}
    \footnotemark[$*$] Bins plotted as upper limits are not counted in the degree of freedom.  \\
    \footnotemark[$\dag$] The maximum-likelihood fit is performed with coherent selection of the PSF1+PSF2+PSF3 data in 0.3--100~GeV.  \\
    \footnotemark[$\ddag$] $E_\mathrm{cut}$ for this component is found to diverge to infinity, so that the curve is reduced to PL. \HILI{In such a case, we also determine the lower limit of $E_\mathrm{cut}$ at a 2$\sigma$ level.}
  \end{tabnote}
\end{table*}

In order to examine the existence of a spectral break for each spatial component, we define the Bayesian Information Criterion (BIC; \cite{Jackson2005}) that compares the PL and BKPL fits: 
\begin{equation}
    BIC_\mathrm{br}=(\chi^2_\mathrm{BKPL}-\chi^2_\mathrm{PL})+(k_\mathrm{BKPL}-k_\mathrm{PL})\mathrm{ln}(n_\mathrm{data}),
\end{equation}
where $\chi^2_\mathrm{BKPL}-\chi^2_\mathrm{PL}$ is the $\chi^2$ difference between PL and BKPL, $k_\mathrm{BKPL}-k_\mathrm{PL}$ is the difference between the numbers of their free parameters (simply equal to 2), and $n_\mathrm{data}$ is the sample size (upper limits are not counted in). \PKHY{A negative $BIC_\mathrm{br}$ value indicates  the preference for a spectral break that is justifiable by the data, while a positive $BIC_\mathrm{br}$ value indicates that BKPL is an overfitted model. To complement such model comparisons, we additionally perform F-tests. The results of comparisons are appended to Table~\ref{PL_BKPL}.} The optimal models determined from BIC and F-tests are overlaid as black dash-dot lines in Figures~\ref{SED_Src-NE},~\ref{SED_Src-NW}~and~\ref{SED_Src-S}.

For Src-NE, we obtain $BIC_\mathrm{br}=-17.6$, indicating that BKPL is significantly preferred over PL. \PKHY{Besides, an F-test yields a statistic of 12.8 for (2,5) d.o.f., implying a p-value of 0.011 for the spectral break of Src-NE.} Below its break energy $E_\mathrm{br}=1.50_{-0.50}^{+0.60}$~GeV, Src-NE follows a nearly flat spectrum with an index $\Gamma_1=2.14\pm0.15$. Above $E_\mathrm{br}$, the Src-NE spectrum exhibits a rather steep index $\Gamma_2=2.72_{-0.03}^{+0.10}$. It is reassuring to see that each of these spectral shape parameters is essentially consistent with that of the likelihood-fit model reported in \citet{Yeung2023}. \PKHY{Alternatively, an ECPL model, with a photon index $\Gamma=2.24\pm0.11$ and an exponential cutoff energy $E_\mathrm{cut}=11.7_{-4.1}^{+9.1}$~GeV, is also preferred over PL (BIC$=-17.1$; F-test statistic = 16.5 for (1,6) d.o.f., giving a p-value of 0.0066 for the exponential cutoff).}

For Src-NW, BKPL is also significantly preferred over PL because of $BIC_\mathrm{br}=-5.2$. \PKHY{This is further supported by an F-statistic of 8.3 for (2,6) d.o.f. which corresponds to a p-value of 0.019 for the spectral break of Src-NW.} Below its $E_\mathrm{br}=0.68_{-0.11}^{+0.32}$~GeV, the spectrum has a  hard index $\Gamma_1=0.74\pm1.30$. Above $E_\mathrm{br}$, the spectrum dramatically softens to $\Gamma_2=2.65\pm0.11$. \PKHY{On the other hand, ECPL is not preferable  (i.e. $E_\mathrm{cut}$ is unconstrained).}

$BIC_\mathrm{br}$ for Src-S is found to have a positive value, showing no evidence for a spectral break. \PKHY{Also, the $\Gamma_1>\Gamma_2$ obtained from the BKPL fitting is beyond the single-origin scenario, in accordance with the theoretical framework of this paper. Moreover, a low F-statistic entails a p-value of 0.46 for the spectral break of Src-S. ECPL is not preferable either.} Hence, the Src-S spectrum is satisfactorily described by a PL model with an index $\Gamma=2.49\pm0.06$.

\PKHY{In our spectral energy distributions, the respective approaches to reconstruct the black dash-dot lines and red dotted lines differ in three aspects: (I) Black lines are obtained from energy-dependent selection of data types (PSF1+PSF2+PSF3  below 2~GeV  and FRONT+BACK  above 2~GeV), while red lines are obtained from coherent selection of the PSF1+PSF2+PSF3 data; (II) Black lines are fitted with the Chi-Square statistics, while red lines are fitted with maximum-likelihood estimates; (III) Black lines are fitted to high-level products (i.e. the flux data points), while red lines are reconstructed from analysing the low-level information (i.e. a \emph{Fermi}-LAT 3D counts map). We stress that all these three factors come into the play for the consistency/discrepancy between a black dash-dot line and a red dotted line. Nonetheless, it is reassuring to note that, for each spectral parameter of each component, the values obtained from the two approaches are consistent with each other within the tolerance of statistical uncertainties.}

\section{Discussion \& Summary}\label{DiscSumm}

In this follow-up work, we confirm that BKPL is a favorable model for both the GeV $\gamma$-ray spectra of Src-NE and Src-NW  (while a spectral turnover is significantly detected for Src-NE but not Src-NW in our previous work; \cite{Yeung2023}). Their break energies are consistent with each other within the tolerance of $1.5\sigma$ uncertainties. The broadband PL index $\Gamma$ of Src-NW is harder than that of Src-NE by $2.8\sigma$, supporting a scenario where the emission site of Src-NW is farther away from SNR G298.6$-$0.0 than that of Src-NE (in terms of deprojected 3D distances). The nearly flat spectrum of Src-NE below its break implies that there is another break hidden at $\lesssim$0.3~GeV, below which the photon index becomes well below 2. The 0.3--100~GeV luminosity of Src-NE is $\sim3.57\times10^{35}$~erg~s$^{-1}$ at 10.1~kpc. This value is well within the range of luminosities for SNRs ($10^{35}-10^{36}$~erg~s$^{-1}$; \cite{Bamba2016}).  No spectral break is significantly detected for Src-S from 0.3~GeV to 100~GeV (in both this work and \cite{Yeung2023}).

We stress that the physical implication of a BKPL $\gamma$-ray spectrum for the upstream cosmic rays interacting with an MC is different from that for the downstream cosmic rays interacting with an SNR shell. If the Src-NE emission is predominantly produced at the SNR shell, then its BKPL spectral shape  would reflect the spectral distribution of the cosmic-ray protons trapped at the SNR shock. If we interpret the Src-NW emission as the delayed signature of the cosmic rays that had escaped from the SNR in the past, its $\gamma$-ray spectral break  would correspond to the minimum energy of the hadronic cosmic rays that have already reached the  MCs.

\PKHY{Furthermore, ECPL also serves as a good-fit model for the Src-NE spectrum, whose exponential cutoff energy is associated with the cosmic-ray acceleration limit of the SNR (e.g. \cite{Ahnen2017}) rather than the progress of cosmic-ray escape. This makes it feasible to describe Src-NE as the emission from the shell region of SNR G298.6$-$0.0.}

We are going to derive the cosmic-ray properties  from the $\gamma$-ray spectra of Src-NE, Src-NW and Src-S, \PKHY{and then we probe the relative contributions of the SNR shell and MC to each spatial component} (\S\ref{CRpheno}). Finally, we are going to compare its high-energy features with some analogical SNRs, and place a  further constraint on the age of SNR G298.6$-$0.0  (\S\ref{AgeEstim}).

\subsection{Distinguishing SNR shell and MC $\gamma$-rays based on cosmic-ray phenomenology} \label{CRpheno}

In the followings, we assume a purely hadronic model where the $\gamma$-rays are generated from proton-proton collisions. Concerning the  population of cosmic-ray protons responsible for each of our targeted $\gamma$-ray sources, we introduce a power-law with a minimum-energy bound and a maximum-energy exponential cutoff: 
\begin{equation}
    \frac{dN_\mathrm{CR}}{dE_\mathrm{CR}} =  \left\{
\begin{array}{ll}
	0\ \mathrm{if}\,E_\mathrm{CR}<E_\mathrm{CR,min}  \\
	N_0(\frac{E_\mathrm{CR}}{E_\mathrm{ref}})^{-\Gamma_\mathrm{CR}}\mathrm{exp}(-\frac{E_\mathrm{CR}}{E_\mathrm{CR,max}})\ \mathrm{if}\,E_\mathrm{CR}\geq E_\mathrm{CR,min} 
\end{array}
\right. ,
\end{equation}
\PKHY{where $E_\mathrm{CR}$ refers to the relativistic energy.} Based on this proton population and the cross-sections of \citet{Kafexhiu2014}, we simulate the pion-decay $\gamma$-ray spectrum. Following the scheme of the Chi-Square test of \S\ref{GRspec}, we iterate the cosmic-ray parameters so as to match the observed and simulated $\gamma$-ray spectra. If the exponential cutoff energy $E_\mathrm{CR,max}$ is unconstrainedly high, we fix $E_\mathrm{CR,max}$ at infinity in the iteration of other parameters. In such a case, we also determine the lower limit of $E_\mathrm{CR,max}$ at a 2$\sigma$ level. The $\gamma$-ray spectra corresponding to the best-fit cosmic-ray models are also overlaid in Figures~\ref{SED_Src-NE},~\ref{SED_Src-NW}~and~\ref{SED_Src-S} for comparison. The fitted cosmic-ray parameters are tabulated in Table~\ref{CRparam}. \PKHY{By executing the Minuit processor MINOS of the Python package ``iminuit", the correlations among parameters are taken into account in computing the uncertainties.}

\begin{table*}
    \caption{Spectral parameters of the cosmic-ray populations responsible for different $\gamma$-ray components.}
    \begin{tabular}{lccc}
\hline            & Src-NE & Src-NW                 & Src-S                   \\
\hline            $E_\mathrm{CR,min}$  (GeV) &  $1.38_{-0.16}^{+0.47}$  \{$1.39_{-0.17}^{+0.65}$\}\footnotemark[$\ddag$]  &  $3.17_{-0.90}^{+1.15}$  &  $1.38_{-0.16}^{+0.56}$ \\
$\Gamma_\mathrm{CR}$       &  $2.57_{-0.21}^{+0.18}$ \{2.81 $\pm$ 0.02\}\footnotemark[$\ddag$]  & 2.81 $\pm$ 0.12 & 2.73 $\pm$ 0.09 \\
$E_\mathrm{CR,max}$  (GeV) &  $240_{-150}^{+240}$  \{$\infty$\}\footnotemark[$\ddag$]  & $\infty$($>$120)\footnotemark[$*$]      & $\infty$($>$750)\footnotemark[$*$]    \\
$\chi^2$/d.o.f.\footnotemark[$\#$]  &   6.5  / 5      \{10.2 / 6\}\footnotemark[$\ddag$]         &     3.6    / 7     &  19.4  / 7                         \\  \hline
\end{tabular}

    \label{CRparam}
  \begin{tabnote}
    \footnotemark[$\#$] Bins plotted as upper limits are not counted in the degree of freedom.  \\
    \footnotemark[$\ddag$] The values inside curly brackets demonstrate a less preferable model for Src-NE that has no exponential cutoff. \\
    \footnotemark[$*$] If the exponential cutoff energy $E_\mathrm{CR,max}$ is unconstrainedly high, we fix $E_\mathrm{CR,max}$ at infinity in the iteration of other parameters. In such a case, we also determine the lower limit of $E_\mathrm{CR,max}$ at a 2$\sigma$ level. \\
  \end{tabnote}
\end{table*}

The cosmic-ray population for Src-NE has a minimum energy $E_\mathrm{CR,min}=1.38_{-0.16}^{+0.47}$~GeV that closely approaches the dynamical threshold (1.22~GeV) for pion production in proton-proton interactions. The exponential cutoff is statistically required ($\Delta\mathrm{BIC}=-1.6$ with respect to a model without an exponential cutoff) and at a rather low energy $E_\mathrm{CR,max}=240_{-150}^{+240}$~GeV,  supporting  an old SNR age ($>$10~kyr; initially put forward by \cite{Yeung2023}). Noticeably, such a spectral softening above tens or hundreds of GeV is  consistent with a radiative shell model that considers the re-acceleration of pre-existing cosmic rays by a radiative shock (e.g., \cite{Uchiyama2010, Lee2015}). The cosmic-ray protons trapped inside an SNR are theoretically expected to follow a nearly flat spectrum \PKHY{($\Gamma_\mathrm{CR}\sim2.0$)} between $E_\mathrm{CR,min}$ and $E_\mathrm{CR,max}$. Whereas, the proton index $\Gamma_\mathrm{CR}=2.57_{-0.21}^{+0.18}$ of Src-NE below $E_\mathrm{CR,max}$ is softer than a flat spectrum by a \PKHY{marginal significance} of $2.7\sigma$. The energy dependence of the cosmic-ray escape from an SNR could result in a difference in proton index between the shell and MC regions \citep{Ohira2010,Caprioli2010,Kawanaka2021}. \PKHY{Considering  these factors as well as the spatial coincidence of Src-NE with SNR G298.6$-$0.0, we argue that the detected $\gamma$-ray emission at Src-NE is dominated by the SNR shell, while the nearer MCs interacting with the SNR also account for a minor  portion that is only  concentrated in a lower energy range.}

\PKHY{For Src-NW, the  position of its point-source morphology adopted in the \emph{Fermi}-LAT flux measurements is  offset from the SNR (Figure~\ref{schematic}), disfavouring the SNR shell as its dominating origin.} The cosmic-ray spectrum responsible for Src-NW starts with a higher $E_\mathrm{CR,min}=3.17_{-0.90}^{+1.15}$~GeV, and follows a power-law of $\Gamma_\mathrm{CR}=2.81\pm0.12$. No exponential cutoff is detected for this cosmic-ray population.  The $E_\mathrm{CR,min}$ ratio of Src-NW to Src-NE is $\gtrsim$2, supporting the interpretation that Src-NW is the delayed $\gamma$-ray emission from the farther MCs which are being collided by protons that had earlier escaped from SNR G298.6$-$0.0, since the cosmic-ray escape develops from higher energies to lower energies gradually \citep{Ptuskin2003, Ptuskin2005}. According to Figure~\ref{schematic}, \PKHY{one MC clump (namely MC-W) is  spatially coincident with Src-NW and at approximately the same distance from us as G298.6$-$0.0 ($\sim$10.1~kpc;  \cite{Yeung2023}). MC-W is separated from the SNR center by a deprojected  distance of $\sim$32~pc. This estimate of distance is  consistent with the cosmic-ray diffusion length $l_\mathrm{dif}\sim\sqrt{2D_\mathrm{ISM}T_\mathrm{dif}}\sim50$~pc (e.g. \cite{Atoyan1995}), assuming a diffusion coefficient $D_\mathrm{ISM}$ of $4\times10^{28}$~cm$^2$~s$^{-1}$ at 1~GeV and a diffusion time $T_\mathrm{dif}$ of 10~kyr.}

We also probe the cosmic-ray energy budget required to account  for the $\gamma$-ray flux of Src-NW. We recall that this SNR--MC system is at $\sim$10.1~kpc from us  \citep{Yeung2023}. The integrated energy flux of Src-NW from 0.3~GeV to 100~GeV is $\sim8.60\times10^{-6}$~MeV~cm$^{-2}$~s$^{-1}$, corresponding to a $\gamma$-ray luminosity of $L_{\gamma}\sim1.05\times10^{41}$~MeV~s$^{-1}$  (about half of that of Src-NE). As indicated in Figure~\ref{schematic}, MC-W (potentially a counterpart of Src-NW) has an angular radius of $\sim4.2^{'}$, corresponding to a physical radius of $R_\mathrm{MC}\sim12.3$~pc. This MC clump has a \PKHY{peak column density of protons\footnote{The number of protons per unit area, integrated along the line of sight through the center of the MC clump.}} of $N_\mathrm{H,col}\sim1.45\times10^{22}$~cm$^{-2}$, derived with a CO-to-H$_2$ conversion factor of $2.0\times10^{20}$~cm$^{-2}$~(K~km~s$^{-1}$)$^{-1}$ \citep{Bolatto2013}. Assuming a spherical geometry, the volume density of protons in this MC clump is $n_\mathrm{H,MC}=N_\mathrm{H,col}/2R_\mathrm{MC}\sim190$~cm$^{-3}$ and its volume is $V_\mathrm{MC}=4\pi{R^3_\mathrm{MC}}/3\sim2.31\times10^{59}$~cm$^3$. The average inelastic cross-section for p-p interaction is $\sigma_\mathrm{pp}\sim3.1\times10^{-26}$~cm$^2$ \citep{Kafexhiu2014}. Cosmic-ray particles travel through space almost at the speed of light $c$. In an inelastic proton-proton collision, only $\eta_\mathrm{p\to\gamma}\sim10\%$ of the cosmic-ray energy is converted to the $\gamma$-ray energy (e.g., \cite{Casanova2022}). We thereby compute the required cosmic-ray energy density at Src-NW to be $u_\mathrm{CR}=L_{\gamma}/n_\mathrm{H,MC}\sigma_\mathrm{pp}c\eta_\mathrm{p\to\gamma}V_\mathrm{MC}\sim26$~eV~cm$^{-3}$, which is much higher than the average energy density of local cosmic-rays ($\approx$1~eV~cm$^{-3}$; e.g., \cite{Cummings2016}). \PKHY{This budget is equivalent to a required cosmic-ray energy of $\varepsilon_\mathrm{CR}=u_\mathrm{CR}V_\mathrm{MC}\sim10^{49}$~erg. This $\varepsilon_\mathrm{CR}$ is affordable by a typical supernova explosion, because it releases canonical energy of $\sim10^{51}$~erg with an efficiency of $\sim$10\% for converting kinetic energy to non-thermal cosmic-ray energy \citep{Ginzburg1964}.} This series of calculations allows us to rule out the Galactic cosmic-ray sea as a dominating origin of the gamma-ray emission at Src-NW and, in turn, strengthens the interpretation that the Src-NW radiation is predominantly originated from the cosmic-rays from SNR G298.6$-$0.0.

The proton population accounting for Src-S also follows a power-law without an exponential cutoff. As demonstrated by Figure~\ref{schematic}, no MC along our line of sight to Src-S is detected at approximately the same distance from us as G298.6$-$0.0. Therefore, the MC clump responsible for the Src-S radiation, if it exists, should be at a deprojected 3D distance of $>$41~pc from the SNR G298.6$-$0.0's center. Theoretically, since Src-S is even farther away from SNR G298.6$-$0.0 than Src-NW,  we would expect the proton spectrum for Src-S to be, at the very least, harder than that for Src-NW. Whereas, we observe the opposite: $E_\mathrm{CR,min}=1.38_{-0.16}^{+0.56}$~GeV for Src-S is lower than that for Src-NW by a factor of $\gtrsim$2 while $\Gamma_\mathrm{CR}=2.73\pm0.09$ for Src-S closely matches that for Src-NW. These findings contradict the assumption that SNR G298.6$-$0.0 is the only cosmic-ray source for Src-S. In other words, additional sources  are strongly required to explain the cosmic-ray and $\gamma$-ray spectra of Src-S, at least for the lower energies. \PKHY{For instance, both  SNR G298.5$-$0.3 (whose eastern edge is adjacent to Src-S) and the Galactic cosmic-ray sea are  candidate origins for Src-S, but the limited studies of MCs along this line of sight make it difficult to estimate their relative contributions. We recall that the BKPL model of Src-S (despite its non-preference by the BIC and F-test) weakly suggests a spectral hardening above $\sim$4.4~GeV, giving a faint hint of the possibility of a double-origin scenario.}

\PKHY{We have hereby established a purely hadronic scenario which could reproduce all $\gamma$-ray emissions in and around SNR G298.6$-$0.0. However, we have not yet examined the feasibility of a leptonic model, which is beyond the scope of this paper.}

\subsection{Comparing SNR G298.6$-$0.0 with analogical SNRs} \label{AgeEstim}

For SNR W28, SNR W44, SNR HESS J1731-347 and SNR HB9, GeV $\gamma$-rays at both shell and MC regions are also detected with \emph{Fermi}-LAT \citep{Cui2018, Peron2020, Cui2019, Oka2022}, making them somewhat analogous to our target SNR G298.6$-$0.0 in this work. We make a comparison among their natures and properties (Table~\ref{CompFourSNR}), and summarise their similarities and differences as follows.

\begin{table*}
    \caption{Comparison among five SNRs whose GeV $\gamma$-rays are detected from both shell and MC regions with \emph{Fermi}-LAT.}
    \begin{tabular}{lccccc}
\hline   SNR            & Age (kyr) & \begin{tabular}[c]{@{}c@{}}Harder-Spectrum\\ Region\end{tabular} & \begin{tabular}[c]{@{}c@{}}Hadronic Model \\ for Shell\end{tabular} & \begin{tabular}[c]{@{}c@{}}Hadronic Model \\ for MC\end{tabular} & References\footnotemark[$*$] \\
\hline   G298.6$-$0.0     & 10--30     & Src-NW (MC)                                                                          & Viable                                                                       & Viable                                                                    & (a), (b)   \\ \hline
W28            & 30--40  & MC                                                                          & Viable                                                                       & Viable                                                                    & (c), (d)        \\
W44            & $\sim$20  & MC                                                                          & Viable                                                                       & Viable                                                                    & (e), (f)        \\ \hline
HESS J1731-347 & 2--6       & Shell                                                                       & Disfavoured                                                                       & Viable                                                                    & (g), (h)        \\
HB9            & $\sim$6.6 & MC                                                                          & Disfavoured                                                                       & Viable                                                                    & (i), (j)       \\ \hline   
\end{tabular}

    \label{CompFourSNR}
  \begin{tabnote}
    \footnotemark[$*$] References: (a) This work; (b) \citet{Yeung2023}; (c) \citet{Cui2018}; (d) \citet{Velazquez2002}; (e) \citet{Peron2020}; (f) \citet{Wolszczan1991}; (g) \citet{Cui2019}; (h) \citet{Cui2016}; (i) \citet{Oka2022}; (j) \citet{Leahy2007}. 
  \end{tabnote}
\end{table*}

The hadronic mechanism (pion-decays in proton-proton collisions) \PKHY{is a viable option for explaining} all of their MC spectra, as well as the shell spectra of G298.6$-$0.0, W28 and W44. On the other hand, the leptonic components dominate the shell spectra of HESS J1731-347 and HB9 because of their relatively young ages ($\lesssim$7~kyr). The dominant leptonic cosmic rays of such a young SNR HESS J1731-347 could also elucidate why its shell spectrum is harder than the hadronic MC emission.

\PKHY{Noteworthily, there are more resemblances among G298.6$-$0.0, W28 and W44. For Src-NE that we argue to be dominated by the G298.6$-$0.0 shell, the nearly flat $\gamma$-ray spectrum below its break resembles the spectra of the W28 shell (Figure~5(a) of \cite{Cui2018}) and the W44 shell (Figure~1 of \cite{Peron2020}). On the other hand, for Src-NW that we interpret as an MC clump interacting with SNR G298.6$-$0.0, the even harder $\gamma$-ray spectrum below its break (despite the large uncertainty of the photon index) is also witnessed for an MC clump interacting with SNR W28 (HESS J1800-240A; Figure~5 of \cite{Cui2018}) and an MC clump interacting with SNR W44 (``NW-Source"; Figure~3 of \cite{Peron2020}). Also, the spectral breaks of these three MC clumps are at very similar photon energies. Thus, we propose that the below-break photon index of an SNR-associated source might serve as an additional indicator for distinguishing MC emission from shell emission.}

The high similarity among the $\gamma$-ray properties of G298.6$-$0.0, W28 and W44, where the latter two are both $\gtrsim$20~kyr old, further supports an old age of $>$10~kyr for G298.6$-$0.0, that is initially estimated by \citet{Yeung2023} based on the $\gamma$-ray spectral break energy of Src-NE.  The exponential cutoff energy of the proton spectrum at the W44 shell was derived to be $71\pm6$~GeV \citep{Peron2020}, that is consistent with or lower than what we derive for the G298.6$-$0.0 shell ($E_\mathrm{CR,max}=240_{-150}^{+240}$~GeV).  For the proton spectrum at the W28 shell, \citet{Cui2018} derived the exponential cutoff energy to be an even lower value of $\sim$12~GeV.  In view of the gradual decline of the cosmic-ray escape energy with the SNR age, G298.6$-$0.0 could possibly be as old as W44, \PKHY{but there is no evidence that it is older than W28}. We thereby constrain the age of SNR G298.6$-$0.0 to be 10--30~kyr, \PKHY{under the assumption of a purely hadronic scenario}.

\section*{Acknowledgments}

The \textit{Fermi} LAT Collaboration acknowledges generous ongoing support
from a number of agencies and institutes that have supported both the
development and the operation of the LAT as well as scientific data
analysis.
These include the National Aeronautics and Space Administration and the
Department of Energy in the United States, the Commissariat \`a l'Energie
Atomique
and the Centre National de la Recherche Scientifique / Institut National de
Physique
Nucl\'eaire et de Physique des Particules in France, the Agenzia Spaziale
Italiana
and the Istituto Nazionale di Fisica Nucleare in Italy, the Ministry of
Education,
Culture, Sports, Science and Technology (MEXT), High Energy Accelerator
Research
Organization (KEK) in Japan, and
the K.~A.~Wallenberg Foundation, the Swedish Research Council and the
Swedish National Space Board in Sweden.
Additional support for science analysis during the operations phase is
gratefully
acknowledged from the Istituto Nazionale di Astrofisica in Italy and the
Centre
National d'\'Etudes Spatiales in France. This work was performed in part
under DOE
Contract DE-AC02-76SF00515.

PKHY thanks the Japan Society for the Promotion of Science (JSPS) fellowship (id. PE21024). This work is also supported in part by Grants-in-Aid for Scientific Research from the Japanese Ministry of Education, Culture, Sports, Science and Technology (MEXT) of Japan, JP19K03908 (AB) and JP23H01211 (AB). S.H.L. acknowledges support by JSPS Grant No. JP19K03913 and the World Premier International Research Center Initiative (WPI), MEXT, Japan. T. M. acknowledges supports by JSPS KAKENHI Grant Number 23H01186. PKHY thanks C.-H. Hsieh and C. Y. Law for very useful discussion.  The authors thank the internal review of the \emph{Fermi} LAT Collaboration, especially the comments from G. Mart\'i-Devesa. The authors also thank the journal review conducted by the editor T. Sako and an anonymous referee.

\appendix

\section{Crosschecks on the flux measurements}
\label{FTBK_PSF123_2GeV}

\begin{figure}
    \centering
    \includegraphics[width=80mm]{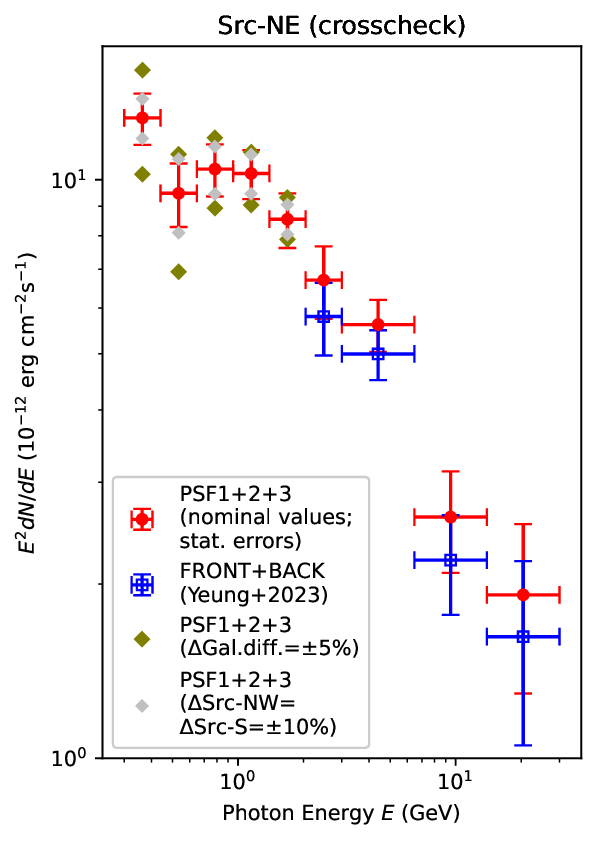}
    \caption{The $>$2~GeV binned spectra of Src-NE (a partial duplicate of Figure~\ref{SED_Src-NE}). Red filled circles are reconstructed with PSF1+PSF2+PSF3 data,   and blue open squares are reconstructed with FRONT+BACK data \citep{Yeung2023}. The error bars plotted here are purely statistical. Dark-yellow diamonds represent the altered fluxes in response to shifting the normalisation of the Galactic diffuse model by $\pm$5\%, while the grey diamonds represent the altered fluxes in response to simultaneously shifting the normalisations of Src-NW and Src-S by $\pm$10\%.}
    \label{apx}
\end{figure}

\PKHY{In order to assess the sufficiency of improvement achieved by PSF1+PSF2+PSF3 data, we test the anticorrelations of the Src-NE spectrum with Src-NW, Src-S and the Galactic diffuse background. Figure~\ref{apx} (a partial duplicate of Figure~\ref{SED_Src-NE}) demonstrates the changes of the Src-NE's $<$2~GeV flux points in response to shifting the normalisations of the Galactic diffuse model, Src-NW and Src-S. It turns out that the integrated energy flux of Src-NE in 0.3--0.65~GeV changes by between +19\% and $-$23\% of the nominal value, and the fractional change of its integrated energy flux in 0.65--2~GeV is only between +10\% and $-$12\%. We, accordingly, judge that the 0.3--2~GeV flux points have been robustly reconstructed with PSF1+PSF2+PSF3 data, and hence a further, tighter cut on the \emph{PSF} event type is not necessary. }

\PKHY{In order to double confirm the $>$2~GeV spectral consistence between PSF1+PSF2+PSF3 data and FRONT+BACK data, we also compare the $>$2~GeV flux points of Src-NE reconstructed with these two selections of event types respectively (Figure~\ref{apx}). It turns out that, for each $\ge$2~GeV bin, the fluxes yielded by the two sets of data are consistent with each other within the tolerance of statistical uncertainties. More importantly, the FRONT+BACK data yields smaller uncertainties due to the higher detection significances. These further justify our  adoption of the FRONT+BACK results  \citep{Yeung2023} above 2~GeV.}


\end{document}